\documentclass{amsart}
\date{\today}
\usepackage{amssymb}
\usepackage{latexsym}

\newcommand{\Z}{{\mathbb Z}}
\newcommand{\R}{{\mathbb R}}

\newtheorem{theorem}{Theorem}

\newtheorem{lemma}{Lemma}[section]

\DeclareMathOperator{\supp}{supp}

\newcounter{smalllist}

\begin{document}
\title[Non-Deterministic Ergodic Potentials]{Ergodic Potentials With a Discontinuous Sampling
Function Are Non-Deterministic}
\author[D.\ Damanik, R.\ Killip]{David Damanik and Rowan Killip}
\address{Department of Mathematics 253--37, California Institute of
Technology, Pasadena, CA 91125, USA, E-mail:
\mbox{damanik@its.caltech.edu}}
\address{Department of Mathematics, University of California, Los Angeles, CA 90055, USA, E-mail:
\mbox{killip@math.ucla.edu}}

\thanks{D.\ D.\ was supported in part by NSF grant DMS--0227289}
\maketitle

\begin{abstract}
We prove absence of absolutely continuous spectrum for discrete
one-dimensional Schr\"odinger operators on the whole line with
certain ergodic potentials, $V_\omega(n) = f(T^n(\omega))$, where
$T$ is an ergodic transformation acting on a space $\Omega$ and $f: \Omega \to \R$.
The key hypothesis, however, is that $f$ is discontinuous.
In particular, we are able to settle a conjecture of Aubry and Jitomirskaya--Mandel'shtam
regarding potentials generated by irrational rotations on the torus.

The proof relies on a theorem of Kotani, which shows that non-deterministic potentials
give rise to operators that have no absolutely continuous spectrum.
\end{abstract}

%%%%%%%%%%%%%%%%%%%%%%%%%%%%%%%%%%%%%%%%%%%%%%%%%%%%%%%%%%%%%%%%%%%%%%%%%%%%%%%%%%%%%%%%%%%
%
%
%                                   Section 1
%
%
%%%%%%%%%%%%%%%%%%%%%%%%%%%%%%%%%%%%%%%%%%%%%%%%%%%%%%%%%%%%%%%%%%%%%%%%%%%%%%%%%%%%%%%%%%%

\section{Introduction}

Consider the discrete quasi-periodic Schr\"odinger operator
$$
[H_\omega \phi](n) = \phi(n+1) + \phi(n-1) + \lambda f(n\alpha + \omega \mod 1) \phi(n)
$$
on $\ell^2(\Z)$ where $\lambda>0$ denotes the coupling constant,
$f: \R/\Z \to \R$ is bounded, $\alpha$ is irrational, and $\omega\in[0,1)$.

When $f$ is continuous, the existence of absolutely continuous spectrum is a difficult
question that has attracted a great deal of attention recently.  In this note, we will
show that for discontinuous $f$, there is none.

The absence of absolutely continuous spectrum for such $f$ was
conjectured by Aubry \cite{a} and by Jitomirskaya and Mandel'shtam
\cite{MZ}.

\begin{theorem}\label{T1}
If $f$ has a single {\rm (}non-removable{\rm )} discontinuity, then for all $\omega\in[0,1)$,
the operator $H_\omega$ has no absolutely continuous spectrum.
\end{theorem}

We will actually prove a far more general result, see Theorem~\ref{T2}.  Irrational
rotations on $\R/\Z$ can be replaced by a very general dynamical system and the restrictions
on $f$ can be significantly relaxed.

In addition to addressing the conjecture described above, this particular example makes a link
to some recent research, which we will now describe.

If $f$ is nice enough, one expects purely absolutely continuous
spectrum for small $|\lambda|$ and purely point spectrum (with
exponentially decaying eigenfunctions) when~$|\lambda|$ is large.
(The latter scenario is usually referred to as Anderson localization.)
Matters are particularly well understood when $f(x) = \cos (2 \pi x)$;
see Jitomirskaya \cite{j} and the references therein. In this case, the operator $H$
has purely absolutely continuous spectrum (for almost all $\alpha$
and $\omega$) when $|\lambda| < 2$ and localization occurs for $|\lambda| > 2$
(again for almost all $\alpha$ and $\omega$).

Under analyticity assumptions on $f$, localization was shown by
Bourgain and Goldstein, \cite{b2,bg}, for $|\lambda|$ large
enough; while for sufficiently small $\lambda$, Bourgain and
Jitomirskaya, \cite{b2,bj}, proved purely absolutely continuous
spectrum.

It is desirable to prove these results under weaker regularity
assumptions on $f$ (see \cite{b} and \cite{k4} for recent
developments in this direction) and, at the same time, explore the
breakdown of these results once $f$ becomes too singular.

It is known that these results definitely do break down when $f$ takes
only finitely many values. In this case, absence of localization
in almost all cases was shown by Delyon and Petritis in \cite{dp}
and absence of absolutely continuous spectrum in all cases is a
result of Kotani \cite{k1}. Notice that these results hold for all
values of the coupling constant~$\lambda$.

We will now describe the setting for our more general result.

Let $T\!:\!\Omega\to\Omega$ be a homeomorphism of a compact metric space $\Omega$
and let $d\mu$ be a probability measure on $\Omega$ with respect to which $T$ is ergodic.
Given a bounded measurable function $f\!:\!\Omega\to\R$, we associate a potential to
each $\omega\in\Omega$ by
$$
V_\omega(n) = f\bigl( T^n(\omega) \bigr) \quad\text{for all $n\in\Z$}.
$$
The corresponding Schr\"odinger operator is denoted by $H_\omega$:
$$
[H_\omega \phi](n) = \phi(n+1) + \phi(n-1) + V_\omega(n) \phi(n), \qquad \phi\in\ell^2(\Z).
$$

To describe the requirements we make on $f$ we need to introduce the notion of an
essential discontinuity.  First, we say that $l\in\R$ is an \textit{essential limit}
of $f$ at $\omega_0$ if there exists a sequence $\{\Omega_k\}$ of sets each of positive measure
such that for any sequence $\{\omega_k\}$ with $\omega_k \in \Omega_k$, both
$\omega_k\to\omega_0$ and $f(\omega_k) \to l$.
If $f$ has more than one essential limit at $\omega_0$, we say that $f$ is \textit{essentially discontinuous}
at this point.

\begin{theorem}\label{T2}
Suppose there is an $\omega_0 \in \Omega$ such that $f$ is essentially discontinuous at $\omega_0$ but is continuous at
all points $T^n \omega_0$, $n < 0$. Then $H_\omega$ has no absolutely continuous spectrum for almost every~$\omega$.
\end{theorem}

\noindent\textit{Remark.} Last and Simon \cite[\S\S 5--6]{ls}
proved the following: if a potential $V$ is the limit (in the
Tychonoff topology) of a sequence of translates of a second
potential, $W$, then the absolutely continuous spectrum of the
operator with potential $V$ contains that of the operator with
potential $W$. This theorem permits one to extend the absence of
absolutely continuous spectrum from almost every $\omega$ to every
$\omega$ in some circumstances. We will use this result in the
proof of Theorem~\ref{T1}.

\medskip

Our strategy will be to revisit a result of Kotani \cite{k1}:
Given an ergodic family of aperiodic potentials that take only
finitely many values, the resulting Schr\"odinger operators have
no absolutely continuous spectrum. The proof of this result rests
on an important consequence of Kotani theory \cite{k3,k2,k1,s}. In
the presence of absolutely continuous spectrum, the potentials are
deterministic in the sense that they are uniquely determined by
their values on a half-line. (The exact meaning of this will be
clarified in the next section.) Kotani proved that when $f$ takes
only finitely many values, the induced potentials are
non-deterministic. We will show that functions $f$ obeying the
assumptions of Theorem~\ref{T2} also give rise to
non-deterministic potentials and so deduce the absence of
absolutely continuous spectrum for the corresponding operators.

%%%%%%%%%%%%%%%%%%%%%%%%%%%%%%%%%%%%%%%%%%%%%%%%%%%%%%%%%%%%%%%%%%%%%%%%%%%%%%%%%%%%%%%%%%%
%
%
%                                   Section 2
%
%
%%%%%%%%%%%%%%%%%%%%%%%%%%%%%%%%%%%%%%%%%%%%%%%%%%%%%%%%%%%%%%%%%%%%%%%%%%%%%%%%%%%%%%%%%%%

\section{A Short Review of Kotani Theory}

Kotani theory (cf.~\cite{k3,k2,k1,s}) establishes a relation
between the absolutely continuous spectrum of an ergodic family of
Schr\"odinger operators and the set of energies at which the
Lyapunov exponent vanishes. In this section, we recall some
specific results from this theory that are necessary for our proof
of Theorem~\ref{T2}. Essentially, we summarize parts of
\cite{k1}.

The setting is as follows: Let $d\nu$ be a measure on the space of
potentials $[-\lambda,\lambda]^{\Z}$ for which the shift map
$[SV](n)=V(n+1)$ is ergodic. Note that $S$ is a homeomorphism with
respect to the Tychonoff topology.

The system $([-\lambda,\lambda]^\Z,S,d\nu)$ gives rise to an ergodic family
of Schr\"odinger operators in $\ell^2(\Z)$,
$$
[H(V) \phi](n) = \phi(n+1) + \phi(n-1) + V(n) \phi(n), \quad V \in
\supp \; d\nu.
$$
For each energy $E$, there is an associated fundamental solution (or transfer matrix):
$$
U(n,E,V) = \begin{pmatrix} E - V(n) & -1 \\ 1 & 0 \end{pmatrix}
\times \cdots \times \begin{pmatrix} E - V(1) & -1 \\ 1 & 0 \end{pmatrix}.
$$
The multiplicative ergodic theorem shows that there is a non-random function $\gamma\!:\!\R\to [0,\infty)$,
called the \textit{Lyapunov exponent}, such that
$$
   \lim_{n \to \infty} \frac1n \log \| U(n,E,V) \| = \gamma(E).
$$
for $\nu$-a.e.\ $V \in [-\lambda,\lambda]^\Z$.  We write $N_{d\nu} = \{ E \in \R : \gamma(E) = 0\}$ for the
set of energies at which the Lyapunov exponent vanishes.

\begin{lemma}[Kotani; see \cite{k1}]\label{lemma1}
If $N_{d\nu}$ has zero Lebesgue measure, then $H(V)$ has no
absolutely continuous spectrum for $\nu$-a.e.\ $V$.
\end{lemma}

Given a potential $V \in [-\lambda,\lambda]^\Z$, we will write $V_\pm$ for its restrictions
to $\Z_+ = \{0,1,2,\ldots\}$ and $\Z_-= \{\ldots,-2,-1\}$, respectively.

\begin{lemma}[Kotani; see \cite{k1}]\label{lemma2}
If $N_{d\nu}$ has positive Lebesgue measure, then each of $V_\pm$
determines $V$ uniquely among  potentials in $\supp(d\nu)$.
\end{lemma}

%%%%%%%%%%%%%%%%%%%%%%%%%%%%%%%%%%%%%%%%%%%%%%%%%%%%%%%%%%%%%%%%%%%%%%%%%%%%%%%%%%%%%%%%%%%
%
%
%                                   Section 3
%
%
%%%%%%%%%%%%%%%%%%%%%%%%%%%%%%%%%%%%%%%%%%%%%%%%%%%%%%%%%%%%%%%%%%%%%%%%%%%%%%%%%%%%%%%%%%%

\section{Absence of Absolutely Continuous Spectrum}

In this section we will prove Theorem~\ref{T2} and then use this
to prove Theorem~\ref{T1}.

Let us fix $\lambda>0$ once and for all such that
$[-\lambda,\lambda]$ contains the range of $f$. As in the
introduction, we associate a potential
$V_\omega\in[-\lambda,\lambda]^\Z$ to each $\omega\in\Omega$ by
$V_\omega(n) = f(T^n(\omega))$.  We write $d\nu$ for the measure
on $[-\lambda,\lambda]^\Z$ induced by the measure $d\mu$ on
$\Omega$.

When $f$ is continuous, the map $\omega\mapsto V_\omega$ is also continuous and so takes
$\supp(\mu)$ onto $\supp(\nu)$. (Under a continuous map, the image of a compact set is compact,
while the inverse image of a closed set is closed.)  When $f$ is not continuous, it is possible for
the support of $d\nu$ to contain potentials that do not correspond to any point $\omega\in\Omega$.
This plays an important role in the proof of Theorem~\ref{T2} which we will now present.

\begin{proof}[Proof of Theorem~\ref{T2}.]
We will show that there are potentials in $\supp(d \nu)$ which
agree on $\Z_-$ but take different values at $n=0$.  By
Lemma~\ref{lemma2} this implies that $N_{d\nu}$ has zero Lebesgue
measure.  Absence of absolutely continuous spectrum for $\nu$-a.e.
potential then follows by Lemma~\ref{lemma1}.

Let $l$ be an essential limit of $f$ at $\omega_0$.  As $f$ is essentially discontinuous at
$\omega_0$, it suffices to construct a potential in $\supp(d\nu)$
that agrees with $V_{\omega_0}$ on $\Z_-$ and takes the value $l$ at $n=0$.

Let $\Omega_k$ be a sequence of sets which exhibit the fact that
$l$ is an essential limit of $f$. Since each has positive
$\mu$-measure, we can find points $\omega_k\in\Omega_k$ so that
$V_{\omega_k}$ is in the support of $\nu$; indeed this is the case
for almost every point in $\Omega_k$.

As $\omega_k \to \omega_0$ and $f$ is continuous at each of the
points $T^n\omega_0$, $n<0$, it follows that $V_{\omega_k}(n)\to
V_{\omega_0}(n)$ for each $n<0$.  Moreover, since $f(\omega_k)$
converges to $l$ we also have $V_{\omega_k}(0)\to l$.  We can
guarantee convergence of $V_{\omega_k}(n)$ for $n>0$ by passing to
a subsequence because $[-\lambda,\lambda]^\Z$ is compact.  Let us
denote this limit potential by~$V$.

As each $V_{\omega_k}$ lies in $\supp(d\nu)$, so does $V$; moreover, $V(0)=l$ and
$V(n)=V_{\omega_0}(n)$ for each $n<0$.  The construction of this potential completes
the proof of the theorem for reasons set forth in the opening paragraphs.
\end{proof}

\begin{proof}[Proof of Theorem~\ref{T1}.]
This result fits into the framework of Theorem~\ref{T2}, were $\Omega = \R / \Z$,
$d\mu$ is Lebesgue measure on the torus $\R / \Z$, and $T$ is the
rotation by $\alpha$, that is, $T (\omega) = \omega + \alpha \mod 1$.  Lastly, we write
$\omega_0$ for the point at which $f$ is discontinuous.

Let us say that $l$ is a limiting value of $f$ at $\omega_0$ if there is
a sequence $\{\omega_k\}$ in $\Omega\setminus\{\omega_0\}$ such that $\omega_k\to\omega_0$
and $f(\omega_k)\to l$.   As $f$ has a non-removable discontinuity at $\omega_0$,
it has more than one limiting value at this point.  Moreover, since $f$ is continuous away
from $\omega_0$, any limiting value is also an essential limit---simply choose each
$\Omega_k$ to be a suitably small interval around $\omega_k$.

This shows that $f$ has an essential discontinuity at $\omega_0$.  As the orbit of $\omega_0$
never returns to this point, $f$ is continuous at each point $T^n\omega_0$, $n\neq0$.
Therefore, Theorem~\ref{T2} is applicable and shows that $H_\omega$ has no absolutely
continuous spectrum for Lebesgue-a.e. $\omega\in[0,1)$.

It remains to show that the absolutely continuous spectrum of
$H_\omega$ is empty for \textit{all} $\omega \in \R / \Z$.
We begin by fixing $\omega_1$ such
that $H_{\omega_1}$ has no absolutely continuous spectrum and such that the orbit
of $\omega_1$ does not meet $\omega_0$; almost all $\omega_1$ have these properties.

Given an arbitrary $\omega\in\R/\Z$, we may choose a sequence of
integers $\{n_i\}$ so that $T^{n_i}(\omega) \to \omega_1$.  As $f$
is continuous on the orbit of $\omega_1$, the potentials
associated to $T^{n_i}(\omega)$ converge pointwise to
$V_{\omega_1}$. By the result of Last and Simon described in the
introduction, the absolutely continuous spectrum cannot shrink
under pointwise approximation using translates of a single
potential. Thus, the operator with potential $V_\omega$ cannot
have absolutely continuous spectrum. This concludes the proof.
\end{proof}

\noindent\textit{Remark.} The proof of Theorem~\ref{T1} shows that for functions $f$
on the torus, every isolated (non-removable) discontinuity is essential.
As a result, the arguments presented extend to show absence of absolutely continuous
spectrum for all $\theta$ when $f$ has finitely many discontinuities.
If $f$ has infinitely many discontinuities, these methods still work in most cases;
Theorem~\ref{T2} describes in detail what is needed from $f$.

\end{document}